\documentclass[showpacs]{revtex4} 
\usepackage{epsfig,amsmath,amssymb,graphicx,upgreek,textcomp}
\usepackage{natbib}
\usepackage[english]{babel} %

\begin{document}

\vspace{0mm}
\title{COMPLEX SCALAR RELATIVISTIC FIELD 
  AS A PROBABILITY AMPLITUDE
} %
\author{Yu.M. Poluektov}
\email{yuripoluektov@kipt.kharkov.ua (y.poluekt52@gmail.com)} %
\affiliation{National Science Center ``Kharkov Institute of Physics and Technology'', 61108 Kharkov, Ukraine} %

\begin{abstract}
A relativistic equation for a neutral complex field as a probability amplitude is proposed. The continuity equation for the probability density is obtained. It is shown that there are two types of excitations of this field, which describe particles with positive energy and different dispersion laws. Based on the Lagrangian formalism, conservation laws are obtained. The transition to secondary quantization is considered. 
\newline%
\end{abstract}

\maketitle 

\section{Introduction}\vspace{-0mm} 
In the works of Klein \cite {Kl}, Gordon \cite {G} and Fock \cite {F1}, \cite {F2} a relativistic equation was obtained for a complex scalar field \(\varphi\left( x \right)\equiv \varphi\left( t,\textbf{x} \right)\) which is applicable to the description of particles with zero spin
\begin{equation} \label{01}
\begin{array}{l}
 \displaystyle\left( \Delta -\frac{1}{c^{2}}\frac{\partial^{2} }{\partial t^{2}}-\mu^{2}\right)\varphi\left( x \right)=0,
\end{array}
\end{equation} 
where \(c\) is the speed of light, \(\mu\equiv mc/\hbar \) is the inverse Compton length, \(m\) is the mass of the particle, \(\Delta \) is Laplace operator. From (1) follows the continuity equation 
\begin{equation} \label{02}
\begin{array}{l}\displaystyle
\frac{\partial \rho}{\partial t}+\nabla \textbf{j}=0,
\end{array}
\end{equation} 
where
\begin{equation} \label{03}
\begin{array}{c}\displaystyle
\rho=\frac{i}{2\mu c}\left( \varphi^{\ast }\frac{\partial \varphi}{\partial t}-\varphi\frac{\partial \varphi^{\ast }}{\partial t} \right), \qquad     \textbf{j}=-\frac{i\hbar }{2m} \left( \varphi^{\ast }\nabla \varphi-\varphi\nabla \varphi^{\ast } \right).
\end{array}
\end{equation} 
In the case of a real field \(\varphi(x)=\varphi(x)^{\ast }\), the continuity equation is absent, and such a field cannot be given a probabilistic interpretation. However, even for a complex field, the quantity \(\rho\) is not positive-definite, and therefore cannot be interpreted as a probability density.  For a free particle, when \(\displaystyle \varphi\sim \exp i\left( \textbf{kx}-\omega t \right)\), the dispersion law  \(\displaystyle \omega^{2}=c^{2}\left( \mu^{2}+k^{2} \right)\) follows from the Klein-Gordon-Fock (KGF) equation, so the energy of a particle can be either positive or negative. Thus, in the relativistic theory of a scalar field there are two fundamental difficulties: 1) physical interpretation of a complex scalar field, 2) the presence in the theory of particles with negative energy. 

The aim of this work is to overcome these difficulties. In this work, an equation for a complex field is obtained, which in the non-relativistic limit transforms into the Schrödinger equation. It is shown that this field describes two types of particles with positive energies and different dispersion laws. The transition to the description of multiparticle systems by the method of secondary quantization is considered.

\section{Relativistic equation with first derivative with respect to time}\vspace{-0mm} %

Unlike the non-relativistic Schrödinger equation, the KGF equation includes a second derivative with respect to time. We obtain a relativistic invariant equation with a first derivative with respect to time. For this purpose, we move from the field \(\varphi\left( x \right)\)  to another field \(\psi\left( x \right)\), which differs from the original field by a time-dependent phase factor 
\begin{equation} \label{04}
\begin{array}{l}\displaystyle
\varphi\left( x \right)=\psi\left( x \right)\exp\left( -i\mu c t \right).
\end{array}
\end{equation}
Substituting (\ref{04}) into the original equation (\ref{01}), we obtain an equation which, along with the second derivative with respect to time, also includes the first derivative 
\begin{equation} \label{05}
\begin{array}{l}\displaystyle
\Delta \psi-\frac{1}{c^{2}}\frac{\partial^{2} \psi}{\partial t^{2}}+2i\frac{\mu}{c}\frac{\partial\psi }{\partial t}=0. 
\end{array}
\end{equation}
In the limit \(c\to \infty \) , from (5) follows the non-relativistic Schrödinger equation 
\begin{equation} \label{06}
\begin{array}{l}\displaystyle
i\hbar \frac{\partial \psi}{\partial t}=-\frac{\hbar ^{2}}{2m}\Delta \psi
\end{array}
\end{equation}
for which the function \(\psi\) admits a Born interpretation as a probability amplitude. It is natural to assume that this interpretation will remain valid for the relativistic case as well. In the original KGF equation (1), spatial coordinates and time enter symmetrically, so that in the new reference system with coordinates   \(x'\equiv \left( t',\textbf{x}' \right)\) this equation will have the same form. When describing in terms of a field \(\psi\left( x \right)\), we break the symmetry between spatial coordinates and time, however, the theory remains relativistically invariant \cite {AB}. Symmetry between coordinates and time is a sufficient condition for relativistic invariance, but is not a necessary condition. Indeed, if we substitute the function 
\begin{equation} \label{07}
\begin{array}{l}\displaystyle
\varphi'\left( x' \right)=\psi'\left( x' \right)\exp \left( -i\mu c t' \right)
\end{array}
\end{equation}
into the KGF equation written in the new coordinate system 
\begin{equation} \label{08}
\begin{array}{l}\displaystyle
\left( \Delta '-\frac{1}{c^{2}}\frac{\partial^{2}}{\partial t'^{2}}-\mu^{2}\right)\varphi'\left( x' \right)=0,
\end{array}
\end{equation}
then we arrive at an equation that coincides in the new coordinate system with equation (\ref{05}): 
\begin{equation} \label{09}
\begin{array}{l}\displaystyle
\Delta '\psi'-\frac{1}{c^{2}}\frac{\partial ^{2}\psi'}{\partial t'^{2}}+2i\frac{\mu}{c}\frac{\partial \psi'}{\partial t'}=0. 
\end{array}
\end{equation}
Note that if for a scalar field \(\varphi\left( x \right)=\varphi'\left( x' \right)\) , then the new function is transformed according to a different law 
\begin{equation} \label{10}
\begin{array}{l}\displaystyle
\psi\left( x \right)\exp\left( -i\mu c t \right)=\psi'\left( x' \right)\exp\left( -i\mu c t' \right) 
\end{array}
\end{equation}
and therefore is not, generally speaking, a scalar with respect to the Lorentz transformation. In a spatial inversion transformation  \(\textbf{x}'=-\textbf{x},\quad t'=t \), the field \(\psi\left( x \right)\)  transforms either as a scalar \(\psi'\left( t,-\textbf{x} \right)=\psi\left( t,\textbf{x} \right)\)  or as a pseudoscalar \(\psi'\left( t,-\textbf{x} \right)=-\psi\left( t,\textbf{x} \right) \). 
\section{Continuity equation for probability density}\vspace{-0mm} %
In order to give the field \(\psi\left( x \right)\)  a physical meaning of probability amplitude, it is necessary to obtain a continuity equation for the probability density. For this purpose, we write equation (\ref{05}) in the form 
\begin{equation} \label{11}
\begin{array}{l}\displaystyle
\frac{\partial \psi}{\partial t}-\frac{i\hbar }{2m}\Delta \psi=-\frac{i\hbar }{2mc^{2}}\frac{\partial^{2}\psi }{\partial t^{2}},
\end{array}
\end{equation}
and we will also introduce dimensionless coordinates and time 
\begin{equation} \label{12}
\begin{array}{l}\displaystyle
\widetilde{x_{i}}\equiv \frac{x_{i}}{L},\quad\tau\equiv \frac{t}{t_{0}},
\end{array}
\end{equation}
where \(L\)  and \(t_{0}\)  is some characteristic length and time. Assuming there is a relationship between these parameters 
\begin{equation} \label{13}
\begin{array}{l}\displaystyle
\frac{\hbar }{2m}\cdot \frac{t_{0}}{L^{2}}=1,
\end{array}
\end{equation}
we arrive at equation (\ref{11}) in dimensionless form 
\begin{equation} \label{14}
\begin{array}{l}\displaystyle
\dot{\psi}-i\widetilde{\Delta}\psi=-i\epsilon\ddot{\psi}.
\end{array}
\end{equation}
Here  \(\dot{\psi}\equiv {\partial \psi /\partial{\tau}}\) and   \(\widetilde{\Delta}\) means differentiation with respect to dimensionless coordinates. A dimensionless parameter is introduced 
\begin{equation} \label{15}
\begin{array}{l}\displaystyle
\epsilon\equiv \left( 2\mu L \right)^{-2}=\left( \frac{\hbar }{2mcL} \right)^{2},
\end{array}
\end{equation}
which is proportional to the square of the ratio of the Compton length to the characteristic length \(L\)  over which the wave function changes. In the non-relativistic limit \( \epsilon=0\) , so that (\ref{14}) becomes the Schrödinger equation. Assuming the parameter \( \epsilon\) to be small, the time derivative on the right-hand side of (\ref{14}) can be expressed through derivatives with respect to coordinates. So, taking into account the terms up to \(\epsilon^{2}\), we find 
\begin{equation} \label{16}
\begin{array}{l}\displaystyle
 \dot{\psi}-i\widetilde{\Delta}=i\left( \epsilon\widetilde{\Delta}^{2}\psi+2\epsilon^{2}\widetilde{\Delta }^{3}\psi\right).
\end{array}
\end{equation}
From this equation follows the continuity equation for the probability density, which in dimensional quantities has the form 
\begin{equation} \label{17}
\begin{array}{l}\displaystyle
\frac{\partial\left| \psi \right|^{2} }{\partial t}+\nabla \left( \textbf{j}_{1}+\textbf{j}_{2}+\textbf{j}_{3}\right)=0,
\end{array}
\end{equation}
where 
\begin{equation} \label{18}
\begin{array}{l}\displaystyle
\textbf{j}_{1}=\frac{i c}{2\mu}\left( \psi\nabla \psi^{\ast }-\psi^{\ast }\nabla \psi \right)
\end{array}
\end{equation}
is the probability flux density in the non-relativistic approximation, 
\begin{equation} \label{19}
\begin{array}{l}\displaystyle
 \textbf{j}_{2}=\frac{ic}{\left( 2\mu \right)^{3}}\left( \psi\nabla \Delta \psi^{\ast }-\psi^{\ast }\nabla \Delta \psi -\nabla \psi\Delta \psi^{\ast }+\nabla \psi^{\ast }\Delta \psi\right),
\end{array}
\end{equation}
\begin{equation} \label{20}
\begin{array}{l}\displaystyle
\textbf{j}_{3}=\frac{2ic}{\left( 2\mu \right)^{5}}\left( \psi\nabla \Delta^{2}\psi^{\ast }-\psi^{\ast }\nabla \Delta^{2}\psi -\nabla \psi\Delta ^{2}\psi^{\ast }+\nabla \psi^{\ast }\Delta ^{2}\psi+\Delta\psi\nabla \Delta\psi^{\ast }-\Delta\psi^{\ast }\nabla \Delta\psi\right)
\end{array}
\end{equation}
these are relativistic corrections of the order of \(\epsilon\)  and   \(\epsilon^{2}\) to the flux density. 

Expressing the second derivative with respect to time in the right-hand side of equation (\ref{14}) through derivatives with respect to coordinates in subsequent orders in powers of the parameter \(\epsilon\), we arrive at a generalization of formula (\ref{16}) 
\begin{equation} \label{21}
\begin{array}{l}\displaystyle
 \dot{\psi}=i\sum_{n=1}^{\infty }2^{2n-1}a_{n}\epsilon^{n-1}\widetilde{\Delta}^{n}\psi,
\end{array}
\end{equation}
where are the numerical coefficients \(a_{n}\equiv \left( 2n-1 \right)!!/\left( 2n \right)!!\) , in particular \(\displaystyle a_{1}=1/2,\quad a_{2}=1/8,\quad a_{3}=1/16\) . In dimensional quantities, formula (\ref{21}) will take the form 
\begin{equation} \label{22}
\begin{array}{l}\displaystyle
\frac{\partial \psi}{\partial t}=i\mu c\sum_{n=1}^{\infty }\frac{a_{n}}{\mu^{2n}}\Delta^{n}\psi=\frac{i\hbar }{m}\sum_{n=1}^{\infty }\frac{a_{n}}{\mu^{2\left( n-1 \right)}}\Delta^{n}\psi.
\end{array}
\end{equation}
From this follows the continuity equation for the probability density in general form 
\begin{equation} \label{23}
\begin{array}{l}\displaystyle
\frac{\partial \left| \psi \right|^{2}}{\partial t}+\nabla \textbf{j}=0,
\end{array}
\end{equation}
where the probability flux density taking into account all orders of magnitude \(\epsilon\)  is determined by the formula 
\begin{equation} \label{24}
\begin{array}{c}\displaystyle
\textbf{j}=\sum_{n=1}^{\infty }\textbf{j}_{n}, 
\vspace {2mm} \\\displaystyle\textbf{j}_{n}=-ic\frac{a_{n}}{\mu^{2n-1}}\sum_{\alpha=1}^{n}\left[ \left( \Delta^{\alpha-1}\psi^{\ast } \right)\left( \nabla \Delta^{n-\alpha} \psi\right)-\left( \Delta^{\alpha-1}\psi \right)\left( \nabla \Delta^{n-\alpha}\psi^{\ast } \right) \right].
\end{array}
\end{equation}
Throughout, it is assumed that the Laplacian to the zeroth power is equal to one \(\Delta^{0}=1\) . Formula (\ref{24}) was obtained in a slightly different way in \cite {P1}. Thus, the theory contains derivatives with respect to higher-order coordinates and is non-local. It seems natural. In contrast to the non-relativistic case, where for a free particle there is only one spatial scale - de Broglie wavelength, in the relativistic case a new scale appears -  Compton wavelength. 

\section{Excitations of a complex field with positive energies }\vspace{-0mm} %
If we substitute the solution in the form of a plane traveling wave \(\psi\sim \exp i\left( \textbf{kx}-\omega t \right)\)  into equation (\ref{24}), we arrive at the dispersion equation
\begin{equation} \label{25} 
\begin{array}{l}\displaystyle
\frac{\omega}{c}=\pm \sqrt{\mu^{2}+k^{2}}-\mu.
\end{array}
\end{equation}
From (\ref{25}) it is clear that equation (\ref{05}), like the original KGF equation, leads to solutions with both positive and negative energies. Thus, although, as has been shown, there is a continuity equation for the probability density for the field \(\psi\left( x \right)\), we still cannot regard it as a field describing particles with positive energies. We will show, using the technique applied in \cite {P2,P3} to the description of Fermi fields, that the function \(\psi\left( x \right)\)  can be represented as the sum of two functions that describe particles with positive energies. 

We expand the general solution of equation (\ref{05}) into a Fourier integral and represent it as the sum of two terms 
\begin{equation} \label{26}
\begin{array}{c}\displaystyle
 \psi\left( x \right)=\int_{-\infty }^{\infty }\psi\left( \textbf{x},t \right)e^{-i\omega t}=\psi^{\left( + \right)}\left( x \right)+\psi^{\left( - \right)}\left( x \right),
 \end{array}
\end{equation}
where in functions 
\begin{equation} \label{27}
\begin{array}{c}\displaystyle
\psi^{\left( \pm  \right)}\left( x \right)=\int_{0}^{\infty }\psi\left( \textbf{x},\pm \omega \right)e^{\mp i\omega t}d\omega
\end{array}
\end{equation}
integration is performed only over non-negative frequencies \(\omega\geqslant 0
\) . Substituting (\ref{26}) into (\ref{05}) we obtain 
\begin{equation} \label{28}
\begin{array}{c}
\displaystyle
\int_{0}^{\infty }\left\lfloor \Delta\psi\left( \textbf{x},\omega \right)+\frac{\omega^{2}}{c^{2}}\psi\left( \textbf{x},\omega \right)+2\mu\frac{\omega}{c}\psi\left( \textbf{x},\omega \right) \right\rfloor e^{-i\omega t}d\omega+\vspace {2mm}\\+\displaystyle\int_{0}^{\infty }\left\lfloor \Delta\psi\left( \textbf{x},-\omega \right)+\frac{\omega^{2}}{c^{2}}\psi\left( \textbf{x},-\omega \right)-2\mu\frac{\omega}{c}\psi\left( \textbf{x},-\omega \right) \right\rfloor e^{i\omega t}d\omega=0.
\end{array}
\end{equation}
Multiplying (\ref{28}) first by \(e^{i\omega' t}\), and then by \(e^{-i\omega' t}\), where \(\omega'>0\), and integrating over time, we arrive at two independent equations: 
\begin{equation} \label{29}
\begin{array}{c}\displaystyle
\left\lfloor \Delta\psi\left( \textbf{x},\omega \right)+\frac{\omega^{2}}{c^{2}}\psi\left( \textbf{x},\omega \right)+2\mu\frac{\omega}{c}\psi\left( \textbf{x},\omega \right) \right\rfloor=0,
\end{array}
\end{equation}
\begin{equation} \label{30}
\begin{array}{c}\displaystyle
\left\lfloor \Delta\psi\left( \textbf{x},-\omega \right)+\frac{\omega^{2}}{c^{2}}\psi\left( \textbf{x},-\omega \right)-2\mu\frac{\omega}{c}\psi\left( \textbf{x},-\omega \right) \right\rfloor =0.
\end{array}
\end{equation}
For plane waves \(\psi\left( \textbf{x},\pm \omega \right)\sim e^{i\textbf{kx}}\) , from (\ref{29}), (\ref{30}) follow the dispersion equations for the fields \(\psi\left( \textbf{x}, \omega \right)\)  and  \(\psi\left( \textbf{x}, -\omega \right)\), respectively 
\begin{equation} \label{31}
\begin{array}{c}\displaystyle
\omega^{\left( + \right)2}_{k}+2\mu c\omega^{\left( + \right)}_{k}-c^{2}k^{2}=0,\quad\displaystyle\omega^{\left( - \right)2}_{k}-2\mu c\omega^{\left( - \right)}_{k}-c^{2}k^{2}=0. 
\end{array}
\end{equation}
Since in (\ref{27}) the integration is performed over positive frequencies, when solving the quadratic equations (\ref{31}), only positive roots should be taken. As a result, we come to the conclusion that the field \(\psi\left( \textbf{x}, \omega \right)\)  describes particles with a dispersion law 
\begin{equation} \label{32}
\begin{array}{l}\displaystyle
\omega^{\left( + \right)}_{k}=c\left( \sqrt{\mu^{2}+k^{2}}-\mu \right),
\end{array}
\end{equation}
which we will call  \((+)\) - particles, and the field \(\psi\left( \textbf{x}, -\omega \right)\)  describes particles with a dispersion law 
\begin{equation} \label{33}
\begin{array}{l}\displaystyle
\omega^{\left( - \right)}_{k}=c\left( \sqrt{\mu^{2}+k^{2}}+\mu \right),
\end{array}
\end{equation}
which we will call \((-)\)  - particles. It's obvious that \(\omega^{\left( \pm  \right)}_{k}=\omega^{\left( \pm  \right)}_{-k}\) . 

For small momentums  \(k^{2}\ll \mu^{2}\) we have 
\begin{equation} \label{34}
\begin{array}{c}\displaystyle
\omega^{\left( + \right)}_{k}\sim c\frac{k^{2}}{2\mu},\quad \omega^{\left( -\right)}_{k}\sim c\left( 2\mu+\frac{k^{2}}{2\mu} \right)
\end{array}
\end{equation}
As we can see, the \((+)\)  - particles are massless, and the \((-)\)   - particles have rest mass \(2m\) . In the ultrarelativistic limit, at \(k^{2}\gg \mu^{2}\)  we have 
\begin{equation} \label{35}
\begin{array}{l}\displaystyle
\omega^{\left( + \right)}_{k}\sim c\left( k-\mu \right),\quad\omega^{\left( - \right)}_{k}\sim c\left( k+\mu \right).
\end{array}
\end{equation}
The speeds of particles of both types are the same:
\begin{equation} \label{36}
\begin{array}{l}\displaystyle
v^{\left( \pm  \right)}_{k}=\frac{\partial\omega^{\left( \pm  \right)}_{k} }{\partial k}=\frac{ck}{\sqrt{\mu^{2}+k^{2}}}.
\end{array}
\end{equation}
Each of the time-dependent functions \(\psi^{\left( + \right)}\left( x \right)\)  and \(\psi^{\left( - \right)}\left( x \right)\)  individually obeys equation (\ref{05}), so that 
\begin{equation} \label{37}
\begin{array}{l}\displaystyle
\Delta\psi^{\left( \pm  \right)}\left( x \right)-\frac{1}{c^{2}}\frac{\partial ^{2}\psi^{\left( \pm  \right)}\left( x \right)}{\partial t^{2}}+2i\frac{\mu}{c}\frac{\partial\psi^{\left( \pm  \right)}\left( x \right) }{\partial t}=0.
\end{array}
\end{equation}
Thus, particles with positive energy are described not by the general solution (\ref{26}), but by two functions \(\psi^{\left( + \right)}\left( x \right)\)  and \(\psi^{\left( - \right)}\left( x \right)\), which should be considered as wave functions of particles of two types. The normalization condition must be met for them 
\begin{equation} \label{38}
\begin{array}{l}\displaystyle
\int_{}^{}d\textbf{x}\left| \psi^{\left( \pm  \right)}\left( x \right) \right|^{2}=1.
\end{array}
\end{equation}
The considered interpretation of a complex scalar field as a field for two types of excitations, one of which is without an energy gap and the other with a gap, is reminiscent of the case of a nonlinear complex field with broken phase symmetry \cite {Gl}. In that case, two types of excitations also arise: massless excitations with a linear dispersion law, and excitations with a finite mass. In the case of a transition to secondary quantization for a nonlinear field, taking into account paired anomalous averages that violate phase symmetry, massless particles acquire a finite mass \cite {P4}. 
\section{Lagrangian formalism. Conservation laws}\vspace{-0mm} %
Equations (\ref{37}) can be obtained from the Euler-Lagrange equations 
\begin{equation} \label{39}
\begin{array}{l}\displaystyle
\frac{\partial }{\partial t}\frac{\partial \Lambda}{\partial ({\partial \psi /\partial{t}})}+\frac{\partial }{\partial x_{i}}\frac{\partial \Lambda}{\partial ({\partial \psi /\partial{x_{i}}})}-\frac{\partial \Lambda}{\partial \psi}=0
\end{array}
\end{equation}
if the density of the Lagrange function is chosen in the form 
\begin{equation} \label{40}
\begin{array}{l}\displaystyle
\frac{2m}{\hbar ^{2}}\Lambda^{(\pm )}=\frac{1}{c^{2}}\frac{\partial\psi^{(\pm )\ast } }{\partial t}\frac{\partial\psi^{(\pm ) } }{\partial t}-\nabla \psi^{(\pm )\ast }\nabla \psi^{(\pm )}+i\frac{\mu}{c}\left(  \psi^{(\pm )\ast  }\frac{\partial\psi^{(\pm )} }{\partial t}- \psi^{(\pm )  }\frac{\partial\psi^{(\pm )\ast } }{\partial t}\right).
\end{array}
\end{equation}
The coefficient in (\ref{40}) is introduced so that the Lagrangian density has the dimension of energy per unit volume. Further in this section we will omit the indices \((\pm )\)  since the obtained relations apply to both fields. 

Let us first consider the consequences of the invariance of the Lagrangian (\ref{40}) with respect to phase transformations 
\begin{equation} \label{41}
\begin{array}{l}\displaystyle
\psi(x)\to \psi'(x)=\psi(x)e^{i\chi},\quad\psi^{\ast }(x)\to \psi^{\ast' }(x)=\psi^{\ast }(x)e^{-i\chi},
\end{array}
\end{equation}
where \(\chi\)  is a real parameter. For infinitely small transformations we have 
\begin{equation} \label{42}
\begin{array}{c}\displaystyle
\psi'=\psi+\delta\psi,\quad\psi^{\ast '}=\psi^{\ast }+\delta\psi^{\ast }\vspace {2mm}\\\delta\psi=i\chi\psi,\quad\delta\psi^{\ast }=-i\chi\psi^{\ast }.
\end{array}
\end{equation}
From the invariance of the Lagrangian and the Euler-Lagrange equations (\ref{39}) follows the continuity equation 
\begin{equation} \label{43}
\begin{array}{l}\displaystyle
\frac{\partial }{\partial t}\left[ \frac{\partial \Lambda}{\partial ({\partial \psi /\partial{t}})}\psi-\frac{\partial \Lambda}{\partial ({\partial \psi^{\ast } /\partial{t}})} \psi^{\ast } \right]+\frac{\partial }{\partial x_{i}}\left[ \frac{\partial \Lambda}{\partial ({\partial \psi /\partial x_{i}})}\psi-\frac{\partial \Lambda}{\partial ({\partial \psi^{\ast } /\partial x_{i}})} \psi^{\ast } \right]=0.
\end{array}
\end{equation}
There is here 
\begin{equation} \label{44}
\begin{array}{l}\displaystyle
\frac{\partial \Lambda}{\partial \left( {\partial \psi /\partial x_{i}} \right)}
=-\frac{\hbar ^{2}}{2m}\frac{\partial \psi^{\ast }}{\partial x_{i}},       \quad\frac{\partial \Lambda}{\partial \left( {\partial \psi^{\ast } /\partial x_{i}} \right)}=-\frac{\hbar ^{2}}{2m}\frac{\partial \psi}{\partial x_{i}},
\end{array}
\end{equation}
\begin{equation} \label{45}
\begin{array}{l}\displaystyle
\frac{\partial \Lambda}{\partial \left( {\partial \psi /\partial{t}} \right)}=\frac{\hbar ^{2}}{2m}\left( \frac{1}{c^{2}}\frac{\partial\psi^{\ast } }{\partial t} +i\frac{\mu}{c}\psi^{\ast }\right)\quad\frac{\partial \Lambda}{\partial \left( {\partial \psi^{\ast } /\partial{t}} \right)}=\frac{\hbar ^{2}}{2m}\left( \frac{1}{c^{2}}\frac{\partial\psi }{\partial t} +i\frac{\mu}{c}\psi\right).
\end{array}
\end{equation}
Taking these relations into account, we obtain the equation
\begin{equation} \label{46}
\begin{array}{l}\displaystyle
\frac{\partial\left| \psi \right|^{2} }{\partial t}-\frac{i}{2\mu c}\frac{\partial }{\partial t}\left( \psi\frac{\partial\psi^{\ast } }{\partial t}-\psi^{\ast }\frac{\partial \psi}{\partial t} \right)+\frac{ic}{2\mu}\frac{\partial }{\partial x_{i}}\left( \psi\frac{\partial \psi^{\ast }}{\partial x_{i}}-\psi^{\ast }\frac{\partial \psi}{\partial x_{i}} \right)=0.
\end{array}
\end{equation}
The last term in (\ref{46}) is the divergence of the flux density in the non-relativistic approximation. As can be shown using formula 
\begin{equation} \label{47}
\begin{array}{l}\displaystyle
\frac{\partial ^{2}\psi}{\partial t^{2}}=-2\mu^{2}c^{2}\sum_{n=2}^{\infty }\frac{a_{n}}{\mu^{2n}}\Delta ^{n}\psi,
\end{array}
\end{equation}
which follows from (\ref{22}) and equation (\ref{37}), the second term in (\ref{46}) is the divergence of the flux density, taking into account relativistic effects. As a result, we again arrived at the continuity equation for the probability density (\ref{23}) as a consequence of the symmetry of the Lagrangian (\ref{40}) with respect to the phase transformations (\ref{41}). 

The Lagrangian is expressed in the same way in terms of field functions and their derivatives in any reference frame 
\begin{equation} \label{48}
\begin{array}{c}\displaystyle
\Lambda\left( \psi(x),\psi^{\ast }(x),\frac{\partial\psi(x), }{\partial t}, \frac{\partial\psi^{\ast }(x), }{\partial t},\frac{\partial\psi(x) }{\partial x_{i}},\frac{\partial\psi^{\ast }(x) }{\partial x_{i}}\right)=\vspace {2mm}\\=\displaystyle\Lambda\left( \psi'(x'),\psi'^{\ast }(x'),\frac{\partial\psi'(x'), }{\partial t'}, \frac{\partial\psi'^{\ast }(x'), }{\partial t'},\frac{\partial\psi'(x') }{\partial x'_{i}},\frac{\partial\psi'^{\ast }(x') }{\partial x'_{i}}\right).
\end{array}
\end{equation}
Let us consider the consequences of the invariance of the Lagrangian (\ref{40}) with respect to small translations in time \(t\longrightarrow t'+\delta t\)  and space \(x_{i}\longrightarrow x'_{i}=x_{i}+\delta x_{i}\) . From (\ref{10}) it follows 
\begin{equation} \label{49}
\begin{array}{c}\displaystyle
\psi'(t,\textbf{x})=\psi(t,\textbf{x})+\delta\psi(t,\textbf{x}),\vspace {4mm}\\\displaystyle\delta\psi(t,\textbf{x})=\left[ i\mu c\psi(t,\textbf{x}) -\frac{\partial \psi(t,\textbf{x}) }{\partial t}\right]\delta t-\frac{\partial \psi(t,\textbf{x})}{\partial x_{i}}\delta x_{i}.
\end{array}
\end{equation}
For small translations 
\begin{equation} \label{50}
\begin{array}{l}\displaystyle
\delta\Lambda+\frac{\partial \Lambda}{\partial t}\delta t+\frac{\partial \Lambda}{\partial x_{i}}\delta x_{i}=0
\end{array}
\end{equation}
where 
\begin{equation} \label{51}
\begin{array}{c}\displaystyle
\delta\Lambda=\Lambda\left( \psi'(x),\psi'^{\ast }(x),\frac{\partial\psi'(x) }{\partial t},\frac{\partial\psi'^{\ast }(x) }{\partial t},\frac{\partial\psi'(x) }{\partial x_{i}},\frac{\partial\psi'^{\ast }(x) }{\partial x_{i}} \right)-\vspace {2mm}\\\displaystyle-\Lambda\left( \psi(x),\psi^{\ast }(x),\frac{\partial\psi(x) }{\partial t},\frac{\partial\psi^{\ast }(x) }{\partial t},\frac{\partial\psi(x) }{\partial x_{i}},\frac{\partial\psi^{\ast }(x) }{\partial x_{i}} \right).
\end{array}
\end{equation}
Taking into account the Euler-Lagrange equations (\ref{39}), we find 
\begin{equation} \label{52}
\begin{array}{c}
\displaystyle
\frac{\partial }{\partial t}\left[ \frac{\partial \Lambda}{{\partial ({\partial \psi /\partial{t}})}}\delta\psi+ \frac{\partial \Lambda}{\partial ({\partial \psi^{\ast } /\partial{t}})}\delta\psi^{\ast } \right]+
\frac{\partial }{\partial x_{i}}\left[ \frac{\partial \Lambda}{{\partial ({\partial \psi /\partial{x_{i}}})}}\delta\psi+ \frac{\partial \Lambda}{\partial ({\partial \psi^{\ast } /\partial{x_{i}}})}\delta\psi^{\ast } \right]+\vspace {2mm}
\\\displaystyle+\frac{\partial \Lambda}{\partial t}\delta t+\frac{\partial \Lambda}{\partial x_{i}}\delta x_{i}=0.
\end{array}
\end{equation}
From here, taking into account (\ref{44}), (\ref{45}), we have 
\begin{equation} \label{53}
\begin{array}{c}\displaystyle
\frac{\hbar ^{2}}{2m}\frac{\partial }{\partial t}\left[ \left( \frac{1}{c^{2}}\frac{\partial \psi^{\ast }}{\partial t} +i\frac{\mu}{c}\psi^{\ast }\right)\delta\psi+\left( \frac{1}{c^{2}}\frac{\partial \psi}{\partial t} +i\frac{\mu}{c}\psi\right)\delta\psi^{\ast }\right]-\vspace {2mm}\\\displaystyle-\frac{\hbar ^{2}}{2m}\frac{\partial }{\partial x_{i}}\left( \frac{\partial \psi^{\ast }}{\partial x_{i}}\delta\psi +\frac{\partial \psi}{\partial x_{i}}\delta\psi^{\ast } \right)+\frac{\partial \Lambda}{\partial t}\delta t+\frac{\partial \Lambda}{\partial x_{i}}\delta x_{i}=0.
\end{array}
\end{equation}
Let us first consider a constant translation in time \(\delta t=const\) , assuming \(\delta x_{i}=0\) . Taking into account (\ref{49}), we arrive at the continuity equation for the energy density 
\begin{equation} \label{54}
\begin{array}{l} \displaystyle
\frac{\partial \text{H}}{\partial t}+\nabla \textbf{j}_{E}=0,
\end{array}
\end{equation}
where Hamiltonian density 
\begin{equation} \label{55}
\begin{array}{c}\displaystyle
\text{H}=\frac{\partial \Lambda}{{\partial ({\partial \psi /\partial{t}})}}({\partial \psi /\partial{t}})+\frac{\partial \Lambda}{{\partial ({\partial \psi^{\ast } /\partial{t}})}}({\partial \psi^{\ast } /\partial{t}})-\Lambda=\vspace {2mm}\\\displaystyle=\frac{\hbar ^{2}}{2m}\left( \frac{1}{c^{2}}\frac{\partial \psi^{\ast }}{\partial t}\frac{\partial \psi}{\partial t} +\frac{\partial \psi^{\ast }}{\partial x_{i}}\frac{\partial \psi}{\partial x_{i}}\right)
\end{array}
\end{equation}
is obviously positive, and the energy flux density is determined by the formula 
\begin{equation} \label{56}
\begin{array}{c}\displaystyle
\textbf{j}_{E}=-\frac{\hbar ^{2}}{2m}\left( \frac{\partial \psi}{\partial t}\nabla \psi^{\ast }+ \frac{\partial \psi^{\ast }}{\partial t}\nabla \psi\right)=\vspace {2mm}\\\displaystyle=-i\hbar c^{2}\sum_{n=1}^{\infty }\frac{a_{n}}{2\mu^{2n}}\left( \Delta^{n}\psi\nabla \psi^{\ast }-\Delta^{n}\psi^{\ast }\nabla \psi \right).
\end{array}
\end{equation}
The first term in this series determines the energy flux density in the non-relativistic approximation 
\begin{equation} \label{57}
\begin{array}{l}\displaystyle
\textbf{j}_{E0}=-\frac{i\hbar ^{3}}{4m^{2}}\left( \Delta\psi\nabla \psi^{\ast }-\Delta\psi^{\ast }\nabla \psi \right).
\end{array}
\end{equation}
Let us now consider a constant translation in space, assuming \(\delta x_{i}=const\)  and \(\delta t=0 \) . In this case, from (\ref{53}) follows the continuity equation for the momentum density 
\begin{equation} \label{58}
\begin{array}{l}\displaystyle
\frac{\partial p_{i}}{\partial t}+\frac{\partial \sigma_{ik}}{\partial x_{k}}=0,
\end{array}
\end{equation}
where
\begin{equation} \label{59}
\begin{array}{c}\displaystyle
p_{i}=-\frac{\partial \Lambda}{{\partial ({\partial \psi /\partial{t}})}}\frac{\partial \psi}{\partial x_{i}}-\frac{\partial \Lambda}{{\partial ({\partial \psi^{\ast } /\partial{t}})}}\frac{\partial \psi^{\ast }}{\partial x_{i}}=\vspace {2mm}\\\displaystyle=-\frac{i\hbar }{2}\left( \psi^{\ast }\frac{\partial \psi}{\partial x_{i}}-\psi\frac{\partial \psi^{\ast }}{\partial x_{i}} \right)+\frac{i\hbar }{2}\sum_{n=1}^{\infty }\frac{a_{n}}{\mu^{2n}}\left( \frac{\partial \psi}{\partial x_{i}}\Delta^{n}\psi^{\ast }-\frac{\partial \psi^{\ast }}{\partial x_{i}}\Delta^{n}\psi\right)
\end{array}
\end{equation}
is 3-momentum density,
\begin{equation} \label{60}
\begin{array}{c}\displaystyle
\sigma_{ik}=\Lambda\delta_{ik}-\frac{\partial \Lambda}{{\partial ({\partial \psi /\partial{x_{i}}})}}\frac{\partial \psi}{\partial x_{i}}-\frac{\partial \Lambda}{{\partial ({\partial \psi^{\ast } /\partial{x_{i}}})}}\frac{\partial \psi^{\ast }}{\partial x_{i}}\vspace {2mm}=\\\displaystyle=\Lambda\delta_{ik}-\frac{\hbar ^{2}}{2m}\left( \frac{\partial \psi}{\partial x_{i}}\frac{\partial \psi^{\ast }}{\partial x_{k}}+\frac{\partial \psi^{\ast }}{\partial x_{i}}\frac{\partial \psi}{\partial x_{k}} \right)  
\end{array}
\end{equation}
is 3-momentum \(i\)  - component flux density. In formula (\ref{60}) 
\begin{equation} \label{61}
\begin{array}{c}\displaystyle
\Lambda=-\frac{\hbar ^{2}}{2m}\left[ \frac{\partial \psi^{\ast }}{\partial x_{i}} \frac{\partial \psi}{\partial x_{i}}+\frac{1}{2}\left( \psi^{\ast } \Delta\psi-\psi \Delta\psi^{\ast }\right)\right]\vspace {2mm}+\\\displaystyle+\frac{\hbar ^{2}}{2m}\left[ \frac{1}{c^{2}} \frac{\partial \psi^{\ast }}{\partial t} \frac{\partial \psi}{\partial t}+\mu^{2}\sum_{n=2}^{\infty }\frac{a_{n}}{\mu^{2n}}\left( \psi\Delta^{n}\psi^{\ast } -\psi^{\ast }\Delta^{n}\psi\right)\right].
\end{array}
\end{equation}
The first term in (\ref{59}) determines the momentum density in the non-relativistic approximation. In this approximation, the 3-momentum density and the probability flux density (\ref{18}) are related by the relation \(\textbf{p}=m\textbf{j}_{1}\). 
\section{Expansion of wave functions on plane waves}\vspace{-0mm} %
We obtain expansions of solutions of equations (\ref{37}) in plane waves, assuming 
\begin{equation} \label{62}
\begin{array}{l}\displaystyle
\psi\left( \textbf{x},\pm \omega \right)=\frac{1}{(2\pi)^{3/2}}\int_{}^{}\psi(\textbf{k,}\pm \omega)e^{i\textbf{kx}}d\textbf{k}.
\end{array}
\end{equation}
Substituting (\ref{62}) into equations (\ref{29}), (\ref{30}), we get 
\begin{equation} \label{63}
\begin{array}{l}\displaystyle
\int_{}^{}\left( -k^{2}+\frac{\omega^{2}}{c^{2}}\pm 2\mu\frac{\omega}{c} \right)\psi(\textbf{k},\pm \omega)e^{i\textbf{kx}}d\textbf{k}=0.
\end{array}
\end{equation}
From this, taking into account \(\omega\ge 0\) , it follows that the condition 
\begin{equation} \label{64}
\begin{array}{l}\displaystyle
\left( \omega-\omega^{(\pm )}_{k} \right)\psi(\textbf{k},\pm \omega)=0
\end{array}
\end{equation}
must be satisfied. Here, as before, \(\omega^{(\pm )}_{k}=c\sqrt{\mu^{2}+k^{2}}\mp c\mu\) . The solution to equation (\ref{64}) can be written as 
\begin{equation} \label{65}
\begin{array}{l}\displaystyle
\psi(\textbf{k},\pm \omega)=\pm \psi^{(\pm )}(\textbf{k})\delta(\omega-\omega^{(\pm )}_{k}),
\end{array}
\end{equation}
where \(\psi^{(\pm )}(\textbf{k})\)  is an arbitrary momentum function that must be determined from additional conditions. Taking into account formula (\ref{65}), we find the expansion of wave functions on plane waves 
\begin{equation} \label{66}
\begin{array}{l}\displaystyle
\psi^{(\pm )}=\pm \frac{1}{(2\pi)^{3/2}}\int_{}^{}d\textbf{k}\psi^{(\pm )}(\pm \textbf{k})e^{\mp i\left( \omega^{(\pm )}_{k}t-\textbf{kx} \right)}.
\end{array}
\end{equation}
The normalization condition 
\begin{equation} \label{67}
\begin{array}{l}\displaystyle
\int_{}^{}d\textbf{x}\left| \psi^{(\pm )}(x) \right|^{2}=\int_{}^{}d\textbf{k}\left| \psi^{(\pm )}(\pm \textbf{k}) \right|^{2}=1
\end{array}
\end{equation}
must be met. 

Taking into account the expansion (\ref{66}), we write the total Hamiltonian in the form 
\begin{equation} \label{68}
\begin{array}{c}\displaystyle
H^{(\pm )}=\int_{}^{}\text{H}^{(\pm )}d\textbf{x}=\frac{\hbar ^{2}}{2mc^{2}}\int_{}^{}\left| \psi^{(\pm )}(\pm \textbf{k}) \right|^{2}\left( \omega^{(\pm )2}_{k}+c^{2}k^{2} \right)d\textbf{k}=\vspace {2mm}\\\displaystyle=\frac{\hbar }{mc}\int_{}^{}\left| \psi^{(\pm )}(\pm \textbf{k}) \right|^{2}\hbar \omega^{(\pm )}_{k}\sqrt{\mu^{2}+k^{2}}d\textbf{k}.
\end{array}
\end{equation}
Here, formulas (\ref{31}) and relations \(c^{2}k^{2}\mp \mu c \omega^{(\pm )}_{k}=c \omega^{(\pm )}_{k}\sqrt{\mu^{2}+k^{2}}\) were taken into account. If we introduce new functions \(a^{(\pm )}(\pm \textbf{k})\) , defining them by the formula 
\begin{equation} \label{69}
\begin{array}{l}\displaystyle
\psi^{(\pm )}(\pm \textbf{k})=\frac{\sqrt{\mu}}{(\mu^{2}+k^{2})^{1/4}}a^{(\pm )}(\pm \textbf{k})
\end{array}
\end{equation}
then the total energy and total momentum \(\textbf{P}^{(\pm )}=\int_{}^{}\textbf{p}^{(\pm )}d\textbf{x}\)  of the fields will take the following form 
\begin{equation} \label{70}
\begin{array}{l}\displaystyle
H^{(\pm )}=\int_{}^{}\hbar \omega^{(\pm )}_{k}a^{(\pm )\ast }(\pm \textbf{k})a^{(\pm ) }(\pm \textbf{k}) d\textbf{k},
\end{array}
\end{equation}
\begin{equation} \label{71}
\begin{array}{l}\displaystyle
\textbf{P}^{(\pm )}=\int_{}^{}\hbar \textbf{k}a^{(\pm)\ast  } (\pm \textbf{k})a^{(\pm ) }(\pm \textbf{k}) d\textbf{k}.
\end{array}
\end{equation}
When obtaining formulas (\ref{70}), (\ref{71}) we took into account that \(\omega^{(\pm )}\pm \mu c =c\sqrt{\mu^{2}+k^{2}}\). 
\section{Transition to the representation of secondary quantization}\vspace{-0mm} %
As shown above, equation (\ref{05}) for the field is equivalent to equations (\ref{37}) for the fields \(\psi^{\left( + \right)}\left( x \right)\) and \(\psi^{\left( - \right)}\left( x \right)\) , which describe particles with positive energies. When describing multiparticle systems, we must consider fields as operators acting on state vectors. In this case, one should proceed from the general solution \(\psi(x)=\psi^{(+)}+\psi^{(-)}\), which, according to (\ref{66}), has the form 
\begin{equation} \label{72}
\begin{array}{l}\displaystyle
\psi(x)=\frac{1}{(2\pi)^{3/2}}\int_{}^{}d\textbf{k}\left[ \psi^{(+)}(\textbf{k})e^{-i\left( \omega^{(+)}_{k}t-\textbf{kx} \right)}-\psi^{(-)}(-\textbf{k})e^{i\left( \omega^{(-)}_{k}t-\textbf{kx} \right)}\right].
\end{array}
\end{equation}
The generalized momentum is defined by the formula 
\begin{equation} \label{73}
\begin{array}{l}\displaystyle
\pi(x)=\frac{\partial \Lambda}{{\partial ({\partial \psi(x) /\partial{t}})}} 
=\frac{\hbar ^{2}}{2mc^{2}}\left[ \frac{\partial \psi^{\ast }(x)}{\partial t}+i\mu c \psi^{\ast }(x)\right].
\end{array}
\end{equation}
Taking into account the expansion (\ref{72}), we find
\begin{equation} \label{74}
\begin{array}{l}\displaystyle
\pi(x)=\frac{i\hbar }{2(2\pi)^{3/2}}\int_{}^{}d\textbf{k}\frac{\sqrt{\mu^{2}+k^{2}}}{\mu}\left[ \psi^{(+)\ast }(\textbf{k})e^{i\left( \omega^{(+)}_{k}t-\textbf{kx} \right)}+\psi^{(-)\ast }(-\textbf{k})e^{-i\left( \omega^{(-)}_{k}t-\textbf{kx} \right)}\right].
\end{array}
\end{equation}
To move to a quantum description of fields, one must require that commutation relations be satisfied between the field operators and the generalized momentum at coinciding times: 
\begin{equation} \label{75}
\begin{array}{c}\displaystyle
\left[ \psi(t,\textbf{x}),\pi(t,\textbf{x}') \right]\equiv \psi(t,\textbf{x})\pi(t,\textbf{x}')-\pi(t,\textbf{x}')\psi(t,\textbf{x})=i\hbar \delta(\textbf{x}-\textbf{x}'),\vspace {2mm}\\\displaystyle\left[ \psi^{\ast }(t,\textbf{x}),\pi^{\ast }(t,\textbf{x}') \right]=i\hbar \delta(\textbf{x}-\textbf{x}').  
\end{array}
\end{equation}
The remaining commutation relations are zero. The quantities \(\psi^{(+)}(\textbf{k})\)  and \(\psi^{(-)}(\textbf{k})\)   should also be considered as operators, and the quantities  \(\psi^{(+)\ast }(\textbf{k})\) and  \(\psi^{(-)\ast }(\textbf{k})\)  as operators that are Hermitian conjugate to them. We will denote complex and Hermitian conjugations in the same way. Similarly to (\ref{69}) we introduce the operators \(a^{(\pm )}(\pm \textbf{k})\) : 
\begin{equation} \label{76}
\begin{array}{l}\displaystyle
\psi^{(\pm )}(\pm \textbf{k})=\frac{\sqrt{\mu}}{(\mu^{2}+k^{2})^{1/4}}a^{(\pm )}(\pm \textbf{k})
\end{array}
\end{equation}
In that case the operators (\ref{72}), (\ref{74}) will take the form 
\begin{equation} \label{77}
\begin{array}{l}\displaystyle
\psi(x)=\frac{\sqrt{\mu}}{(2\pi)^{3/2}}\int_{}^{}d\textbf{k}\left( \mu^{2}+k^{2} \right)^{-1/4}\left[ a^{(+)}(\textbf{k})e^{-i\left( \omega^{(+)}_{k}t-\textbf{kx} \right)}-a^{(-)}(-\textbf{k})e^{i\left( \omega^{(-)}_{k}t-\textbf{kx} \right)}\right],
\end{array}
\end{equation}
\begin{equation} \label{78}
\begin{array}{l}\displaystyle
\pi(x)=\frac{i\hbar }{2(2\pi)^{3/2}\sqrt{\mu}}\int_{}^{}d\textbf{k}\left( \mu^{2}+k^{2} \right)^{1/4}\left[ a^{(+)\ast }(\textbf{k})e^{i\left( \omega^{(+)}_{k}t-\textbf{kx} \right)}+a^{(-)\ast }(-\textbf{k})e^{-i\left( \omega^{(-)}_{k}t-\textbf{kx} \right)}\right].
\end{array}
\end{equation}
The commutation relations for field operators (\ref{75}) will be satisfied if the following commutation relations for operators \(a^{(\pm )}(\pm \textbf{k})\)  are satisfied: 
\begin{equation} \label{79}
\begin{array}{c}\displaystyle
\left[ a^{(+)}(\textbf{k}), a^{(+)\ast }(\textbf{k}')\right]=\delta\left( \textbf{k}- \textbf{k}'\right),\quad\left[ a^{(-)}(-\textbf{k}), a^{(-)\ast }(-\textbf{k}')\right]=-\delta\left( \textbf{k}- \textbf{k}'\right),\vspace {2mm}\\\displaystyle\left[ a^{(+)}(\textbf{k}), a^{(-)\ast }(-\textbf{k}')\right]=\left[ a^{(-)}(-\textbf{k}), a^{(+)\ast }(\textbf{k}')\right]=0.
\end{array}
\end{equation}
As we can see, the commutation relation for operators \(a^{(+)}(\textbf{k})\)  has a standard form for Bose operators, and the commutator for operators \(a^{(-)}(-\textbf{k})\)  differs in sign. Therefore, it is convenient to introduce the notation 
\begin{equation} \label{80}
\begin{array}{l}\displaystyle
a^{(-)}(-\textbf{k})\equiv b^{(-)\ast }(\textbf{k})
\end{array}
\end{equation}
Taking this denotation into account, the commutation relations will take the form 
\begin{equation} \label{81}
\begin{array}{c}\displaystyle
\left[ a^{(+)}(\textbf{k}), a^{(+)\ast }(\textbf{k}')\right]=\delta\left( \textbf{k}- \textbf{k}'\right),\quad\left[ b^{(-)}(\textbf{k}), b^{(-)\ast }(\textbf{k}')\right]=\delta\left( \textbf{k}- \textbf{k}'\right),\vspace {2mm}\\\displaystyle\left[ a^{(+)}(\textbf{k}), b^{(-) }(\textbf{k}')\right]=\left[ b^{(-)\ast }(-\textbf{k}), a^{(+)\ast }(\textbf{k}')\right]=0.
\end{array}
\end{equation}
When calculating the total energy operator, the Hamiltonian density operator (\ref{55}) should be taken in a form symmetrized with respect to the operators  \(\psi\) and  \(\psi^{\ast }\) 
\begin{equation} \label{82}
\begin{array}{l}\displaystyle
\text{H}=\frac{\hbar ^{2}}{4m}\left( \frac{1}{c^{2}}\frac{\partial \psi^{\ast }}{\partial t}\frac{\partial \psi}{\partial t} +\frac{1}{c^{2}}\frac{\partial \psi}{\partial t}\frac{\partial \psi^{\ast }}{\partial t}+\frac{\partial \psi^{\ast }}{\partial x_{i}}\frac{\partial \psi}{\partial x_{i}}+\frac{\partial \psi}{\partial x_{i}}\frac{\partial \psi^{\ast }}{\partial x_{i}}\right).
\end{array}
\end{equation}
This allows us to take into account the observed effects caused by zero-point field fluctuations. As a result, we obtain the following expression for the operator of the total energy of a multiparticle system  \((+)\) and \((-)\)  particles 
\begin{equation} \label{83}
\begin{array}{c}\displaystyle
H=\frac{1}{2}\int_{}^{}d\textbf{k}[ \hbar \omega^{(+)}_{k}(a^{(+)\ast }(\textbf{k})a^{(+) }(\textbf{k})+a^{(+) }(\textbf{k})a^{(+)\ast }(\textbf{k}))+\vspace {2mm}\\+\hbar \omega^{(-)}_{k}(b^{(-)\ast }(\textbf{k})b^{(-) }(\textbf{k})+b^{(-) }(\textbf{k})b^{(-)\ast }(\textbf{k})) ].
\end{array}
\end{equation}
Note that the cross terms of the type  \(\displaystyle a^{(+)\ast }(\textbf{k})a^{(-)}(-\textbf{k})\) have dropped out in (\ref{83}) due to the relation \(\displaystyle \omega^{(+)}_{k}\omega^{(-)}_{k}=c^{2}k^{2}\) . 
Similarly, when calculating the total momentum operator, the momentum density (\ref{59}) should be taken in symmetrized form 
\begin{equation} \label{84}
\begin{array}{l}\displaystyle
\textbf{p}=-\frac{1}{2}\left( \pi\nabla \psi+\nabla \psi\cdot \pi+\pi^{\ast }\nabla \psi^{\ast } +\nabla \psi^{\ast } \cdot \pi^{\ast }\right).
\end{array}
\end{equation}
As a result, we obtain the total momentum operator in the form 
\begin{equation} \label{85}
\begin{array}{c}\displaystyle
\textbf{P}=\frac{\hbar }{2}\int_{}^{}d\textbf{k}\textbf{k}[ \hbar \omega^{(+)}_{k}(a^{(+)\ast }(\textbf{k})a^{(+) }(\textbf{k})+a^{(+) }(\textbf{k})a^{(+)\ast }(\textbf{k}))+\vspace {2mm} \\\displaystyle+\hbar \omega^{(-)}_{k}(b^{(-)\ast }(\textbf{k})b^{(-) }(\textbf{k})+b^{(-) }(\textbf{k})b^{(-)\ast }(\textbf{k})) ].
\end{array}
\end{equation}
In this case, the cross terms of the type \(\displaystyle a^{(+)\ast }(\textbf{k})a^{(-)}(-\textbf{k})\)  drop out during integration. 

The vector of the ground  state (vacuum) is determined by the relations 
\begin{equation} \label{86}
\begin{array}{l}\displaystyle
a^{(+)}({\bf k})|0\rangle=0,\quad b^{(-)}({\bf k})|0\rangle=0.
\end{array}
\end{equation}
In a vacuum state we have zero-point energy 
\begin{equation} \label{87}
\begin{array}{l}\displaystyle
E=\big\langle 0\big|H\big|0\big\rangle=\frac{V}{2(2\pi)^{3}}\int_{}^{}d\textbf{k}\left( \hbar \omega^{(+)}_{k}+\hbar \omega^{(+-)}_{k} \right)
\end{array}
\end{equation}
and the average value of the momentum \(\big\langle 0\big|\textbf{P}\big|0\big\rangle=0\). We took into account that \(\delta(0)=V/(2\pi)^{3}\), \(V\) is volume. If we do not consider the effects determined by zero-point oscillations, then the energy (\ref{83}) and momentum (\ref{85}) operators can be written in a normally ordered form 
\begin{equation} \label{88}
\begin{array}{l}\displaystyle
H=\int_{}^{}d\textbf{k}[ \hbar \omega^{(+)}_{k}a^{(+)\ast }(\textbf{k})a^{(+) }(\textbf{k})+\hbar \omega^{(-)}_{k}b^{(-)\ast }(\textbf{k})b^{(-) }(\textbf{k})],
\end{array}
\end{equation}
\begin{equation} \label{89}
\begin{array}{l}\displaystyle
\textbf{P}=\hbar \int_{}^{}d\textbf{k}\textbf{k}[ a^{(+)\ast }(\textbf{k})a^{(+) }(\textbf{k})+b^{(-)\ast }(\textbf{k})b^{(-) }(\textbf{k})].
\end{array}
\end{equation}
The state vectors of one particle with momentum \(\hbar \textbf{k}\)  are determined by the relations
\begin{equation} \label{90}
\begin{array}{c}\displaystyle
|1^{(+)}_{k}\rangle\equiv a^{(+)\ast }({\bf k})|0\rangle,\quad |1^{(-)}_{k}\rangle\equiv b^{(-)\ast }({\bf k})|0\rangle 
\end{array}
\end{equation}
and are eigenvectors of the energy (\ref{88}) and momentum (\ref{89}) operators: 
\begin{equation} \label{91}
\begin{array}{c}\displaystyle
H|1^{(+)}_{k}\rangle=\hbar \omega^{(+)}_{k} |1^{(+)}_{k}\rangle ,\quad H|1^{(-)}_{k}\rangle=\hbar \omega^{(-)}_{k} |1^{(-)}_{k}\rangle,\vspace {2mm}\\\displaystyle\textbf{P}|1^{(+)}_{k}\rangle=\hbar \textbf{k} |1^{(+)}_{k}\rangle ,\quad \textbf{P}|1^{(-)}_{k}\rangle=\hbar \textbf{k} |1^{(-)}_{k}\rangle.
\end{array}
\end{equation}
The operators of the number particles of both types are defined by the formulas 
\begin{equation} \label{92}
\begin{array}{c}\displaystyle
N^{(+)}=\int_{}^{}d\textbf{k}a^{(+)\ast }(\textbf{k})a^{(+) }(\textbf{k}),\quad N^{(-)}=\int_{}^{}d\textbf{k}b^{(-)\ast }(\textbf{k})b^{(-) }(\textbf{k}).
\end{array}
\end{equation}
We also give a formula for the commutator between the field and the generalized impulse at non-coinciding times 
\begin{equation} \label{93}
\begin{array}{l}\displaystyle
\left[ \psi(t,\textbf{x}),\pi(t',\textbf{x}') \right]=i\hbar \Delta(t-t',\textbf{x}-\textbf{x}'), 
\end{array}
\end{equation}
where 
\begin{equation} \label{94}
\begin{array}{l}\displaystyle
\Delta(t-t',\textbf{x}-\textbf{x}')=\frac{1}{2(2\pi)^{3}}\int_{}^{}d\textbf{k}\left[ e^{-i\omega^{(+)}_{k}(t-t')} +e^{i\omega^{(-)}_{k}(t-t')}\right]e^{i\textbf{k}\left( \textbf{x}-\textbf{x}' \right)}.
\end{array}
\end{equation}
\section{Conclusion}\vspace{-0mm} %
In the proposed theory introduces a complex function \(\psi\left( x \right)\) , that differs from the scalar function \(\varphi\left( x \right)\)  satisfying the KGF equation by a time-dependent phase factor. For the new function, an equation containing the first derivative with respect to time is obtained. In the non-relativistic limit, this equation becomes the Schrödinger equation. It is shown that the introduced new complex function \(\psi\left( x \right)
\)  can be interpreted as a probability amplitude. The continuity equation for the probability density is obtained. The general solution of the equation for \(\psi\left( x \right)\)  can be represented as the sum of two functions, describing particles with positive energies and different dispersion laws. One type of particle is massless, while the other has a finite rest mass. It can be assumed that such a pair of scalar particles are  \(\pi^{0}\) - meson and heavier \(K^{0}\)  - meson. The mass of a free particle with zero rest mass can appear due to nonlinear effects \cite {P4}. The proposed theory contains derivatives with respect to high-order coordinates and is non-local. The description of a multiparticle system in the transition to a secondary quantization representation is considered. 



\end{document}